\documentclass[aps,prb,preprint,superscriptaddress,showpacs,floatfix]{revtex4}
\usepackage{amsmath,amssymb}
\usepackage{blindtext}
\usepackage{graphicx} 
\begin{document}

\title{Spin-independent v-representability of Wigner crystal oscillations in one-dimensional Hubbard chains: The role of spin-charge separation}
 
\author{Daniel Vieira}
\affiliation{Departamento de F\' isica,
Universidade do Estado de Santa Catarina, Joinville, 89219-710 SC, Brazil}
\date{\today}

\newcommand{\be}{\begin{equation}}
\newcommand{\ee}{\end{equation}}
\newcommand{\bea}{\begin{eqnarray}} 
\newcommand{\eea}{\end{eqnarray}} 
\newcommand{\bi}{\bibitem} 

\renewcommand{\r}{({\bf r})}
\newcommand{\rp}{({\bf r'})}

\newcommand{\ua}{\uparrow}
\newcommand{\da}{\downarrow}
\newcommand{\la}{\langle}
\newcommand{\ra}{\rangle}
\newcommand{\dg}{\dagger}

\begin{abstract}
Electrons in one-dimension display the unusual property of separating their spin and charge into two independent entities: The first, which derive from uncharged spin$-1/2$ electrons, can  travel at different velocities when compared with the second, built from charged spinless electrons. Predicted theoretically in the early sixties, the spin-charge separation has attracted renewed attention since the first evidences of experimental observation, with usual mentions as a possible explanation for high-temperature superconductivity. In one-dimensional (1D) model systems, the spin-charge separation leads the frequencies of Friedel oscillations to suffer a $2k_F\rightarrow4k_F$ crossover, mainly when dealing with strong correlations, where they are  referred to as Wigner crystal oscillations. In non-magnetized systems, the current density functionals which are applied to the 1D Hubbard model are not seen to reproduce this crossover, referring to a more fundamental question: Are the Wigner crystal oscillations  in 1D systems non-interacting v-representable? Or, is there a spin-independent Kohn-Sham potential which is able to  yield spin-charge separation? Finding an appropriate answer to both questions is our main task here. By means of exact and  DMRG solutions, as well as, a new approach of exchange-correlation potential, we show the answer to be positive. Specifically, the v-representable $4k_F$ oscillations emerge from attractive interactions mediated by positively charged spinless holes -- the holons -- as an additional contribution to the repulsive on-site Hubbard interaction.
\end{abstract}

\maketitle

\section{Introduction and background}   

Friedel oscillations are oscillations of the electron density that appear around inhomogeneities and were first described by J. Friedel.\cite{friedel} With renewed attention due to new possibilities of experimental realization, Friedel oscillations have emerged as a useful  laboratory in the study of many-body systems,\cite{friedelapplications1,friedelapplications2} in particular those described by the one-dimensional (1D) Hubbard chains. Contrary to higher dimensional systems, which use to be described by the Fermi-liquid theory, the 1D metals assume non-Fermi-liquid behavior, and belong to a special class of Tomonaga-Luttinger liquids.\cite{luttinger1,luttinger2} 

 In one dimension, and in second-quantized notation,
the Hubbard model\cite{hubbard} (HM) is defined as
\be
\label{hubbard}
\hat{H} = -t \sum_{j, \sigma}^L \left(c_{j
\sigma}^{\dagger}c_{j+1, \sigma} + \textrm{H.c.} \right) +
U\sum_{j}^L c_{j \uparrow}^{\dagger}c_{j \uparrow} c_{j
\downarrow}^{\dagger}c_{j \downarrow} ,
\ee
where $L$ is the number of sites, $t$ the amplitude for hopping
between neighboring sites, and $U$ the local (on-site) interaction acting on site $j$.   
Occupation of each site is
limited to two particles, necessarily of opposite spin. 

\begin{figure*}
\centering
\begin{minipage}[b]{0.5\linewidth}
\includegraphics[scale=0.32]{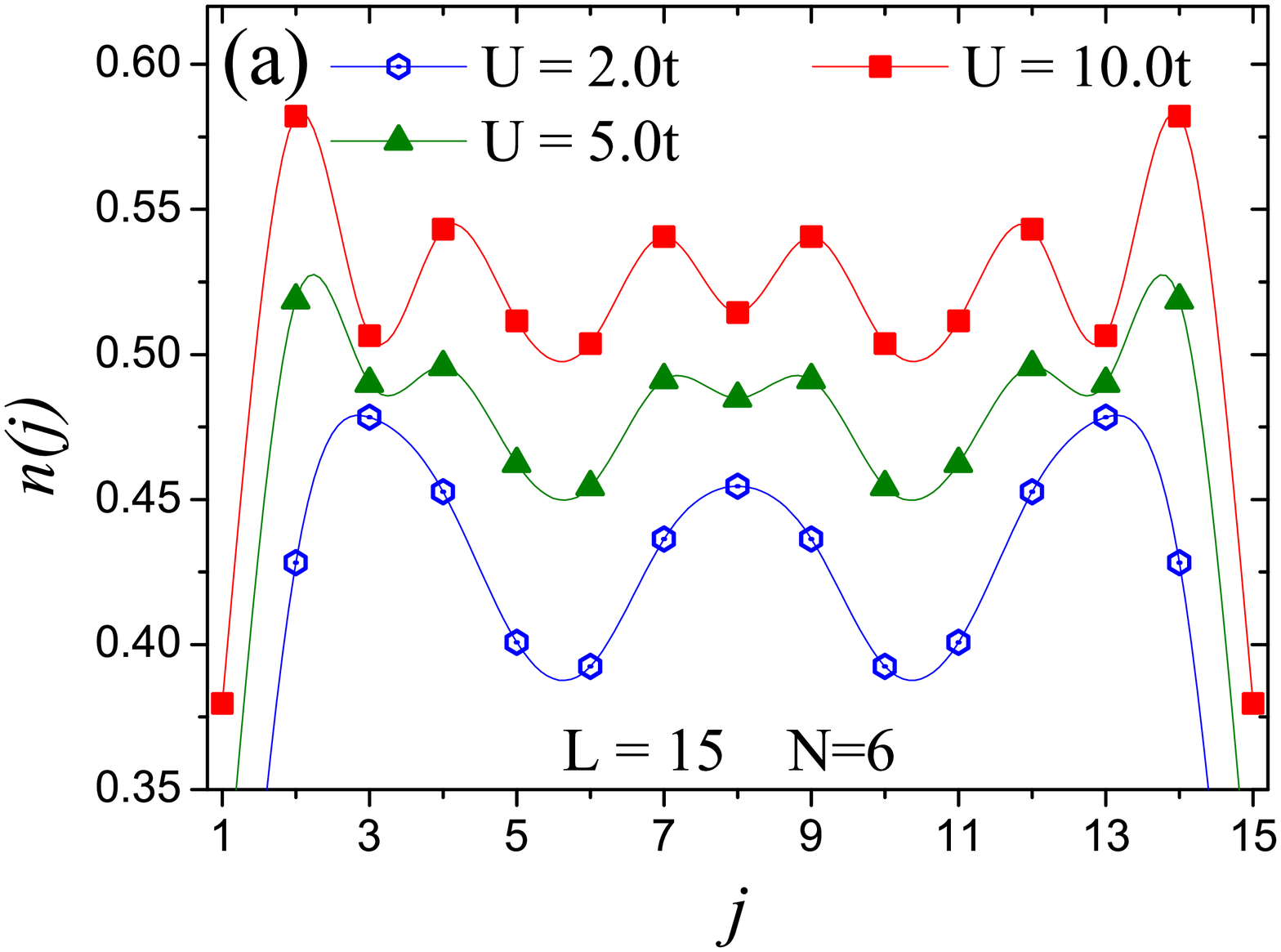}
\end{minipage}\hfill
\begin{minipage}[b]{0.5\linewidth}
\includegraphics[scale=0.32]{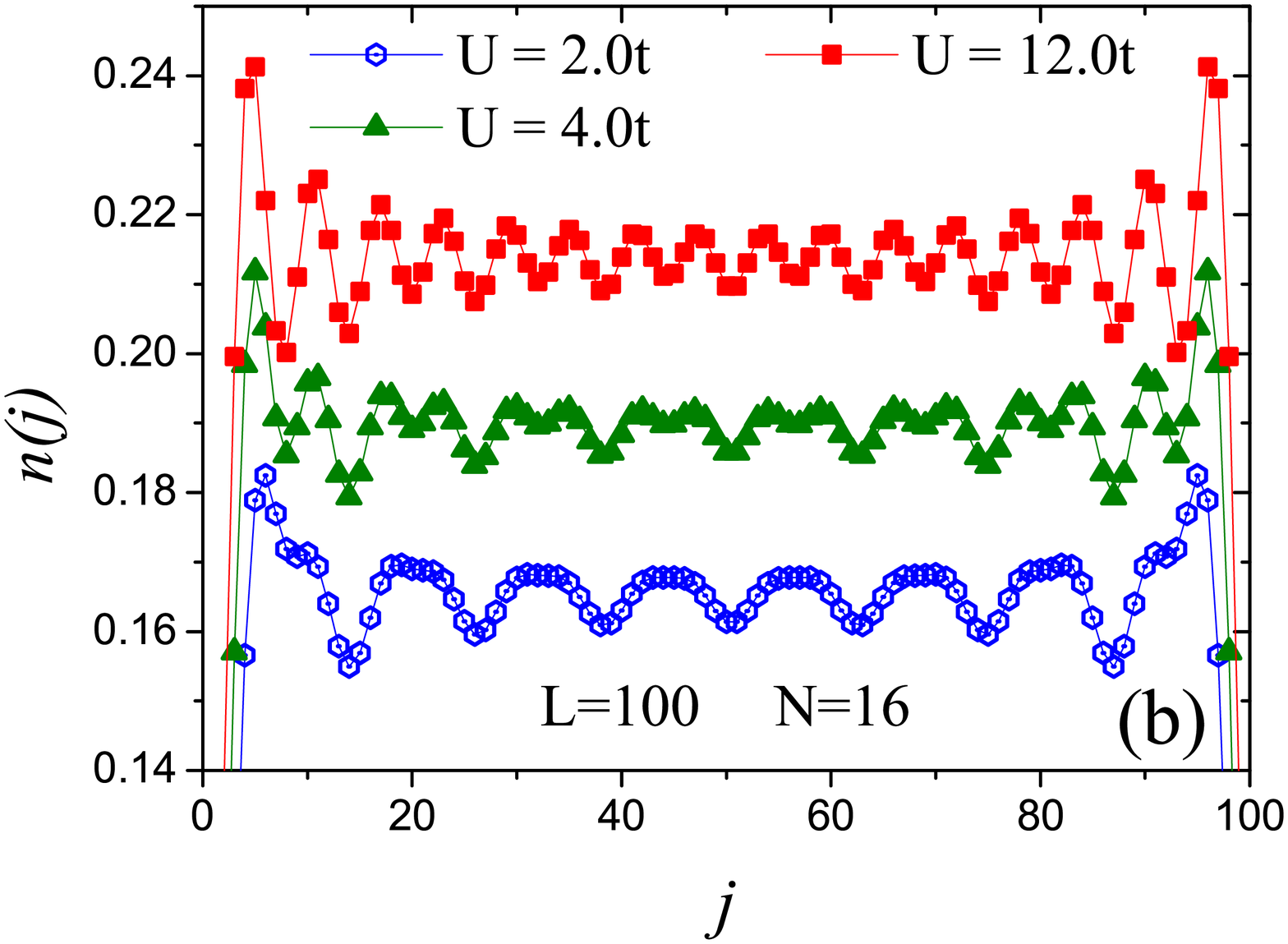}
\end{minipage}
\caption{
\label{fig1} (Color online) Panel (a): Density profiles from Lanczos exact diagonalization, with $L=15$ sites and $N=6$ electrons. Panel (b): the same as before, but from DMRG calculations with $L=100$ sites and $N=16$ electrons. The average densities were added to constant values in order to clarify the visualization.}
\end{figure*}

The density profiles of open Hubbard chains display
Friedel-like oscillations, as it can be seen in Fig.~\ref{fig1}, obtained by means of Lanczos exact diagonalization and DMRG calculations. We note a clear change in the frequencies of oscillation upon variation of the local Hubbard interaction $U$. The increase of $U$ is accompanied by an increase in the frequency, whose value is known to pass from $2k_F$ at $U/t=0$ to $4k_F$ in the limit of $U/t \rightarrow \infty$, corresponding to the transition from three to six positive peaks in Fig.~\ref{fig1}(a) [or from eight to sixteen in Fig.~\ref{fig1}(b)], where $2k_F = \pi (N+1)/(L+1)$, $N$ is the number of electrons and $k_F$ is the Fermi wave vector.\cite{wignerprb,white,noack} Specifically, each positive peak in the density profiles is splitted into two new ones. The resulting $4k_F$ Friedel oscillations, usually referred to as Wigner crystal oscillations,\cite{wignerprb,wigner} are connected with the spin-charge separation effect, where the one-dimensional confinement yields strongly interacting electrons which can break their spin and charge into two separated quasiparticles,  the first built from \mbox{uncharged} spin$-1/2$ and the second from charged spinless electrons.\cite{giamarchi,holes} This mechanism, with recent evidences of experimental observation,\cite{spinchargenature,spinchargeexp,spinchargenature2} is usually mentioned as a possible explanation for high-temperature superconductivity.\cite{super,super2,super3} 

In the limit of $U/t=0$, the density profiles $n(j)$ of Friedel oscillations are exactly described by  
\be
\label{eqfriedel1} 
n^0(j) = \frac{N+1}{(L+1)} -  \frac{\sin(2k_F j)}{ (L+1)\sin \left (\frac{\pi j}{L+1} \right)},
\ee 
whereas, in the limit of $U/t\rightarrow \infty$, the Wigner crystal profiles are given by
\be
\label{eqfriedel2} 
n^\infty(j) = \frac{N+1/2}{(L+1)} -  \frac{\sin\left[(4k_F  - \frac{\pi}{L+1})j \right]}{ 2(L+1)\sin \left (\frac{\pi j}{L+1} \right)},
\ee
which is equivalent to the $U/t= 0$ limit of spinless electrons. 

There have been many attempts to describe the $2k_F\rightarrow4k_F$ crossover by means of a local-density approximation (LDA), or its spin-dependent version (LSDA), which includes to consider magnetized situations with $N_\uparrow \neq N_\downarrow$,\cite{vieira}
and the break of spin symmetry by means of a vanishing magnetic field when $N_\uparrow = N_\downarrow$.\cite{xianlong0,xianlong1,xianlong2} Generalized gradient approximations (GGAs)\cite{mariana} and self-interaction corrections (SICs)\cite{vieira2} have also been considered, with no success for non-magnetized systems. The question we intend to discuss here is: Is it possible, in a non-magnetized system, to reproduce all ranges of frequencies via a standard Kohn-Sham (KS) density-functional theory (DFT) calculation? In other words, we face the v-representability question: Is it possible that the strongly interacting $4k_F$ regime -- and the spin-charge separation -- cannot be yielded by a spin-independent KS potential?

\subsection{The concept of v-representability}

The Hohenberg-Kohn theorems\cite{parryang} state that the total energy of a many-body system is a functional of density, usually written as
\begin{equation}
\label{eq1}
E[n] = T[n] + V_{ee}[n] + V[n],
\end{equation}
standing for the kinetic, interaction and external potential terms, respectively. The exact expression for the ground-state kinetic energy can be written as\cite{parryang}
\begin{equation}
\label{eq2}
T[n] =  \sum_{\sigma=\uparrow,\downarrow} \sum_{k}^{N_\sigma}\  f_{k, \sigma}\ \langle \psi_{k, \sigma}| \hat{t}| \psi_{k, \sigma} \rangle,
\end{equation}
where $\hat{t}$, $\psi_{k, \sigma}$ and $f_{k, \sigma}$ are the kinetic energy operator, spin orbitals and their occupation numbers, respectively. The Pauli exclusion principle requires that $0\leq f_{k, \sigma} \leq 1$. The total electron density is given by
\begin{equation}
\label{eq3}
n\r = \sum_{\sigma=\uparrow,\downarrow} \sum_{k}^{N_\sigma}\  f_{k, \sigma}\ |\psi_{k, \sigma} \r|^2,
\end{equation}
under the constraint of $\sum_{\sigma=\uparrow,\downarrow} \sum_{k}^{N_\sigma} f_{k, \sigma}={N_\uparrow}+N_{\downarrow}=N$. For {\it interacting} systems, with fractional occupation numbers for $f_{k, \sigma}$, there are an infinite number of possibilities in Eqs. (\ref{eq2}) and (\ref{eq3}).
Kohn and Sham considered $f_{k, \sigma}=1$ for the first $N_\sigma$ orbitals and $f_{k, \sigma}=0$ for the rest, that is, they approximated $T[n]$  by the {\it non-interacting} kinetic energy, $T_S[n]$, exactly described by these choices of occupation numbers. In practice,  $V_{ee}[n]$ is also approximated, by the classical Hartree expression, $E_H[n]$, and all corrections to both approximations are considered in the exchange-correlation (XC) term, $E_{xc}[n]$. 

The use of $T_S[n]$,  however, leads to a restriction on the density: It needs to be non-interacting v-representable, that is, there must exist a non-interacting ground-state with the density $n\r$. In other words, there must exist a XC potential $v_{xc}[n]\r = \delta E_{xc}[n]/ \delta n\r $ whose ground-state, in a non-interacting KS calculation, leads to the correct interacting density profiles. There are examples in the literature of densities which are not non-interacting v-representable,\cite{englisch} and a common procedure to circumvent this is to allow fractional occupation numbers for $f_{k, \sigma}$. The v-representability is not a conceptual problem to the Hohenberg-Kohn theorems, since it comes from the choice of $T_S[n]$ for the KS approach of DFT. We can mention alternative approaches, as the orbital-free DFT,\cite{ofdft} which aims to approximate $T[n]$ as an explicit functional of the density, and does not make use of the KS equations. 

\section{Results}

In order to prove the v-representability, we can $(a)$ use accurate solutions of 1D Hubbard chains, such as the DMRG, and invert the KS equation to find the accurate XC potential\cite{xcpotential} which yields the accurate densities; or $(b)$ find an XC potential, or propose a new one, which yields the correct oscillations. In either case, if the potential exists, this will be a {\it numerical} (and sufficient) proof that the $2k_F\rightarrow4k_F$ crossover is non-interacting v-representable. We here will go through these two options.

\subsection{Accurate exchange-correlation potentials} 

\begin{figure*}
\centering
\begin{minipage}[b]{0.5\linewidth}
\includegraphics[scale=0.32]{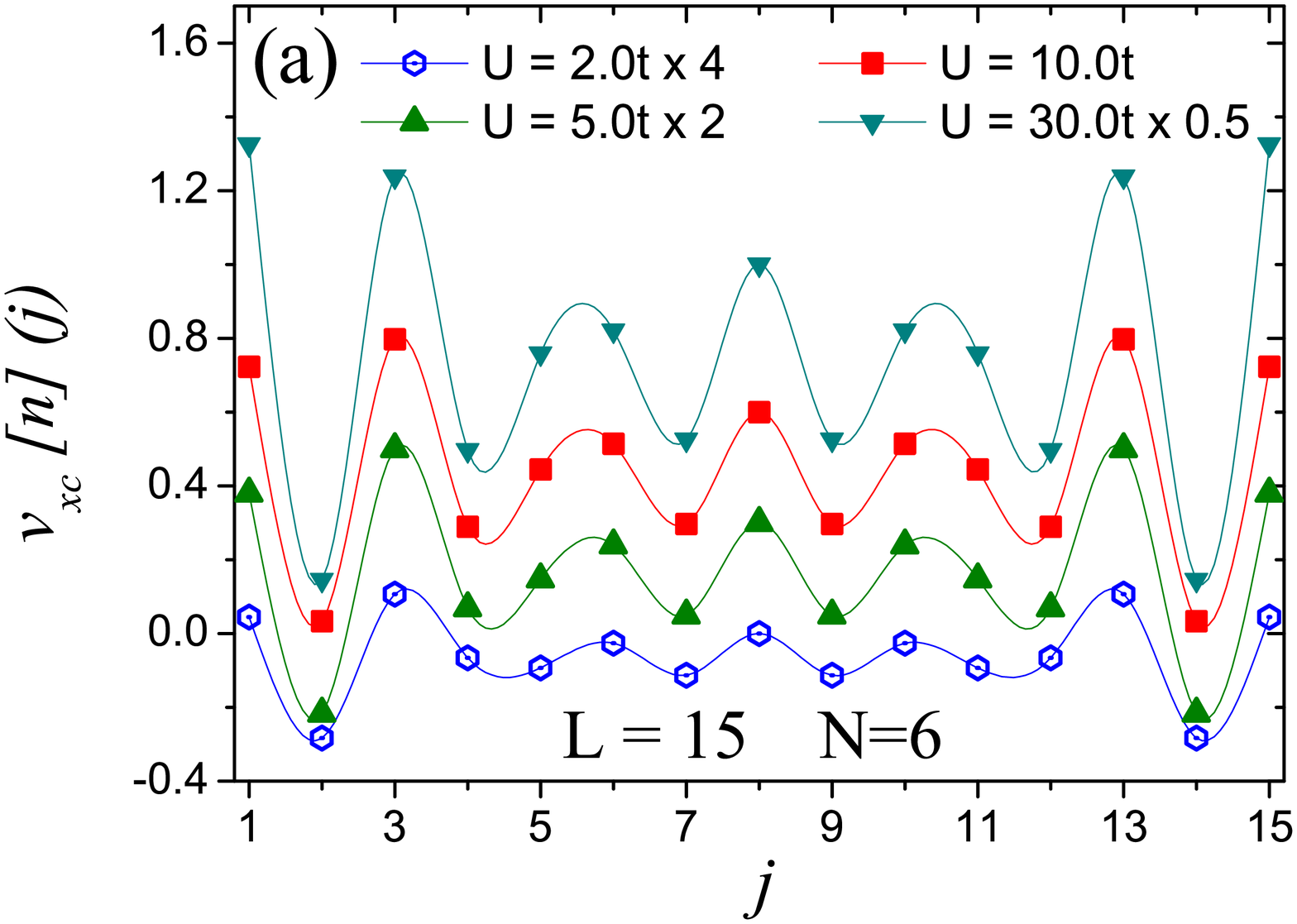}
\end{minipage}\hfill
\begin{minipage}[b]{0.5\linewidth}
\includegraphics[scale=0.32]{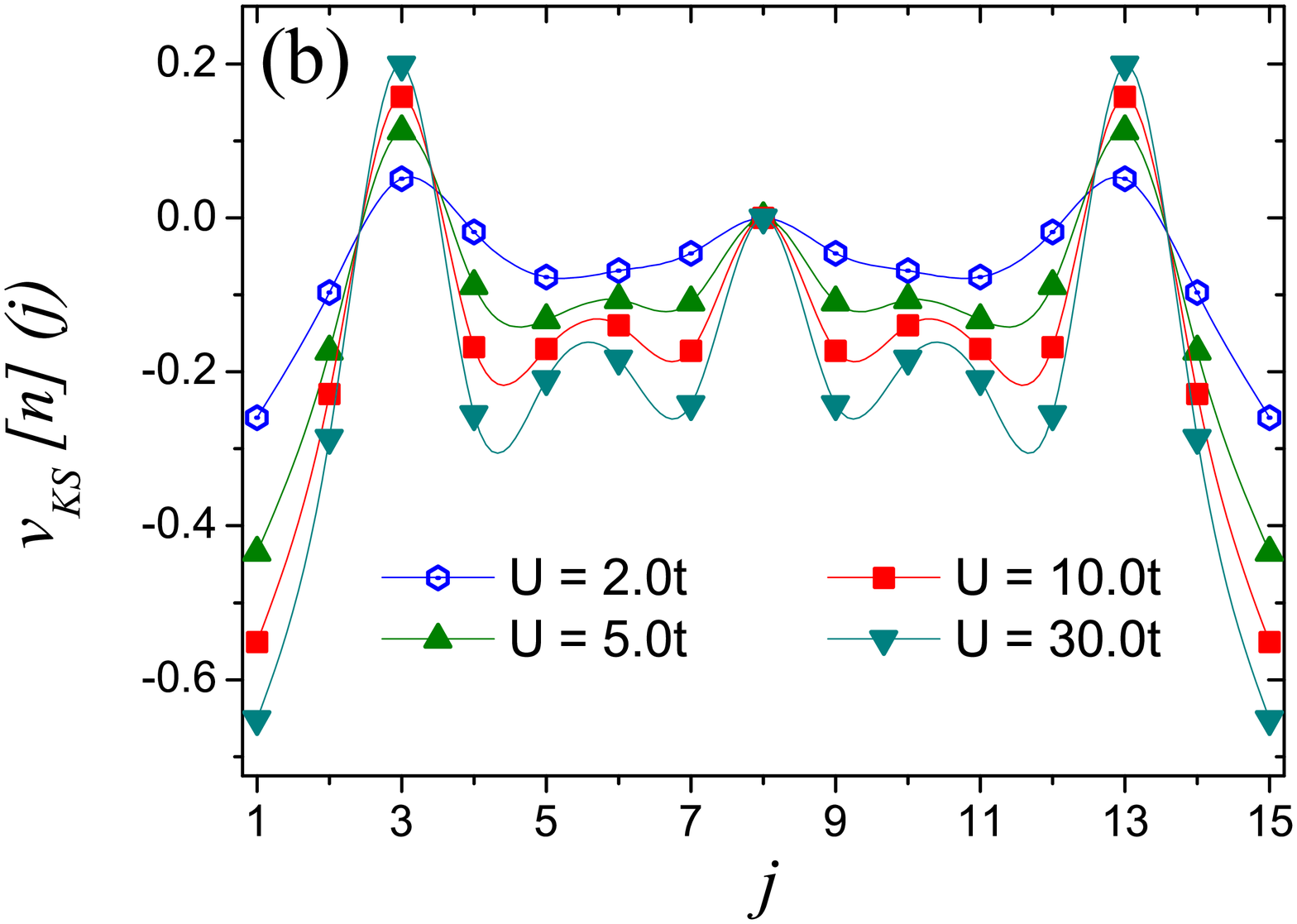}
\end{minipage}
\begin{minipage}[b]{0.5\linewidth}
\includegraphics[scale=0.32]{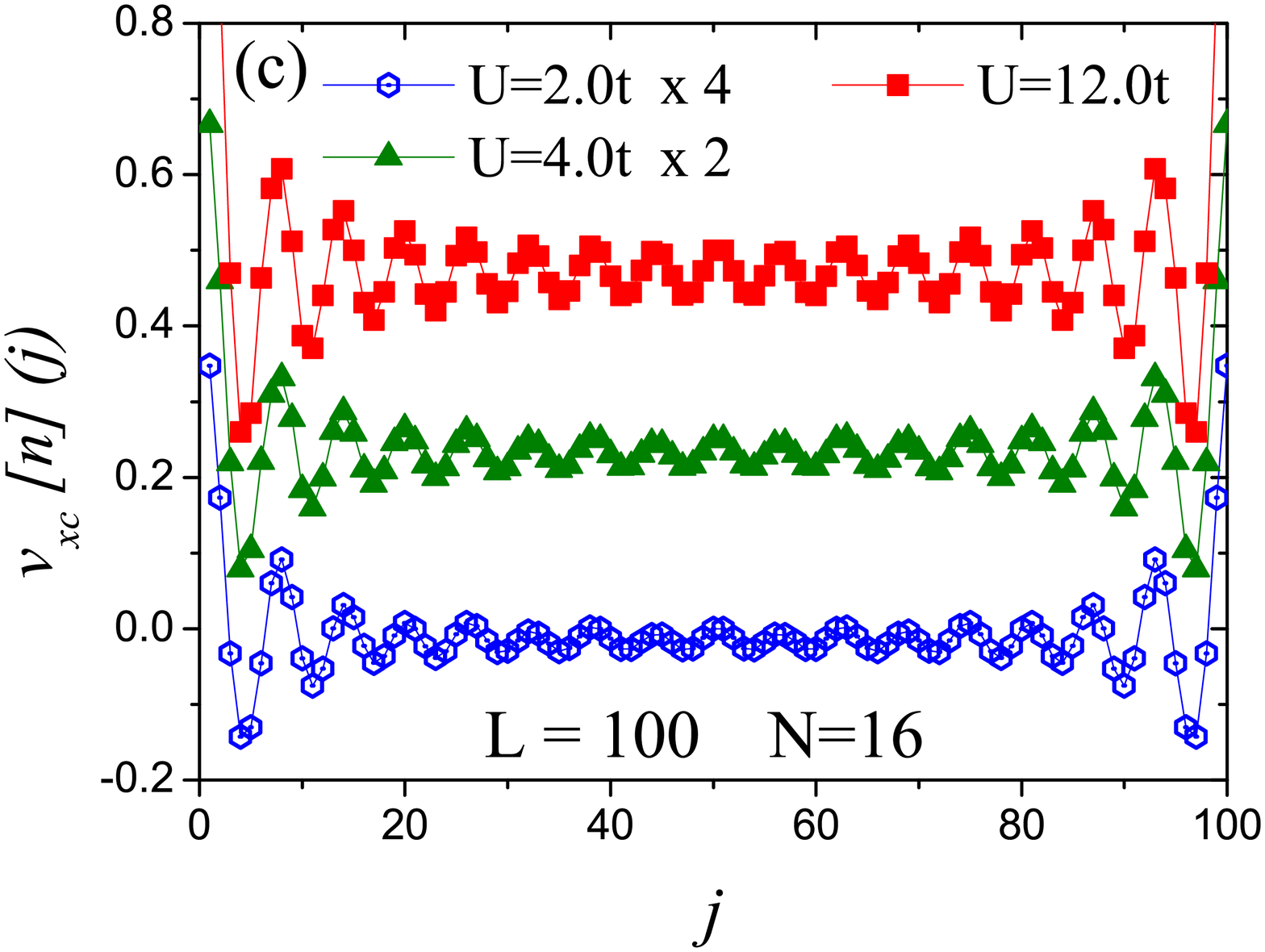}
\end{minipage}\hfill
\begin{minipage}[b]{0.5\linewidth}
\includegraphics[scale=0.32]{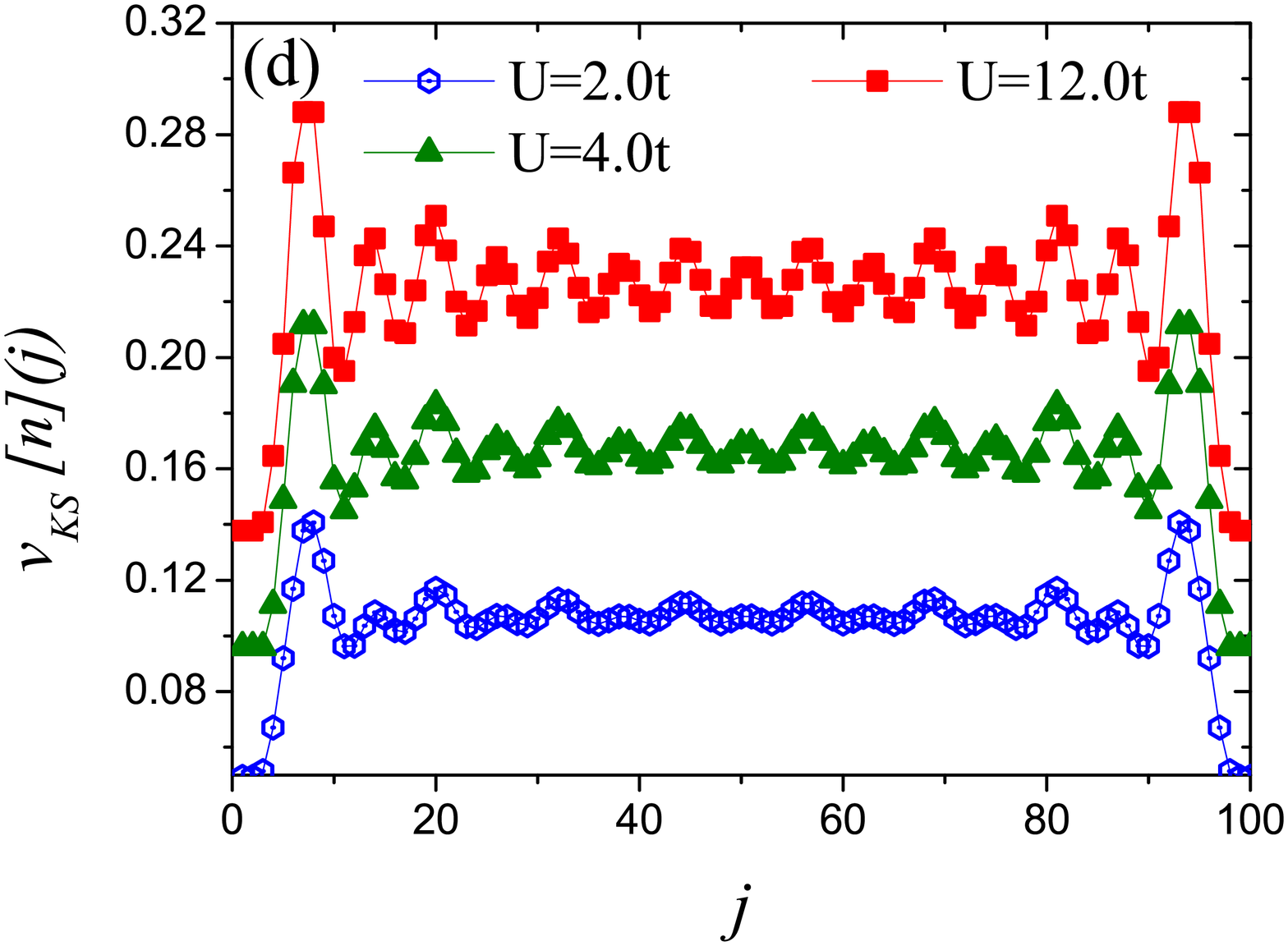} 
\end{minipage}
\begin{minipage}[b]{0.5\linewidth}
\includegraphics[scale=0.32]{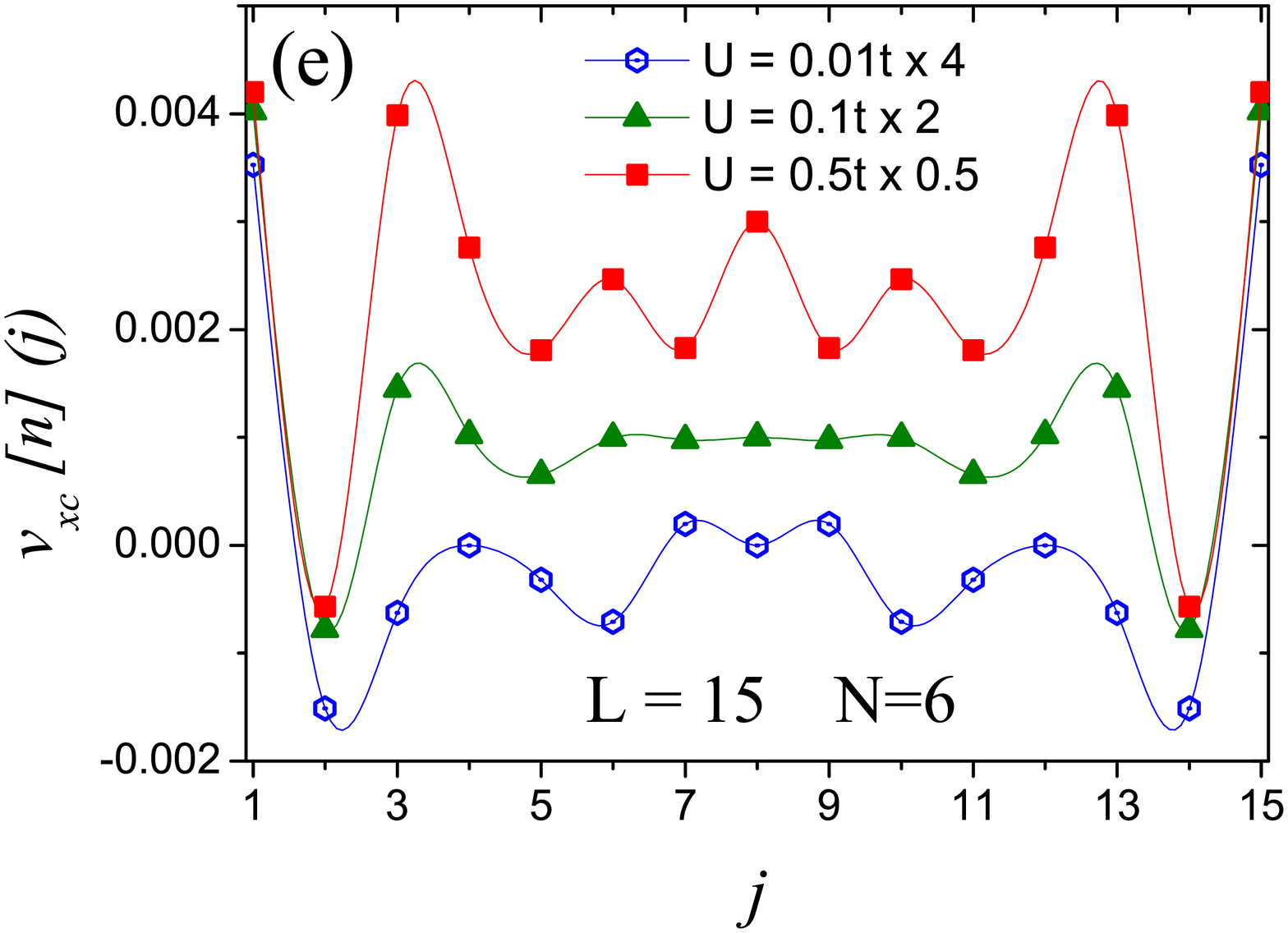}
\end{minipage}\hfill
\begin{minipage}[b]{0.5\linewidth}
\includegraphics[scale=0.32]{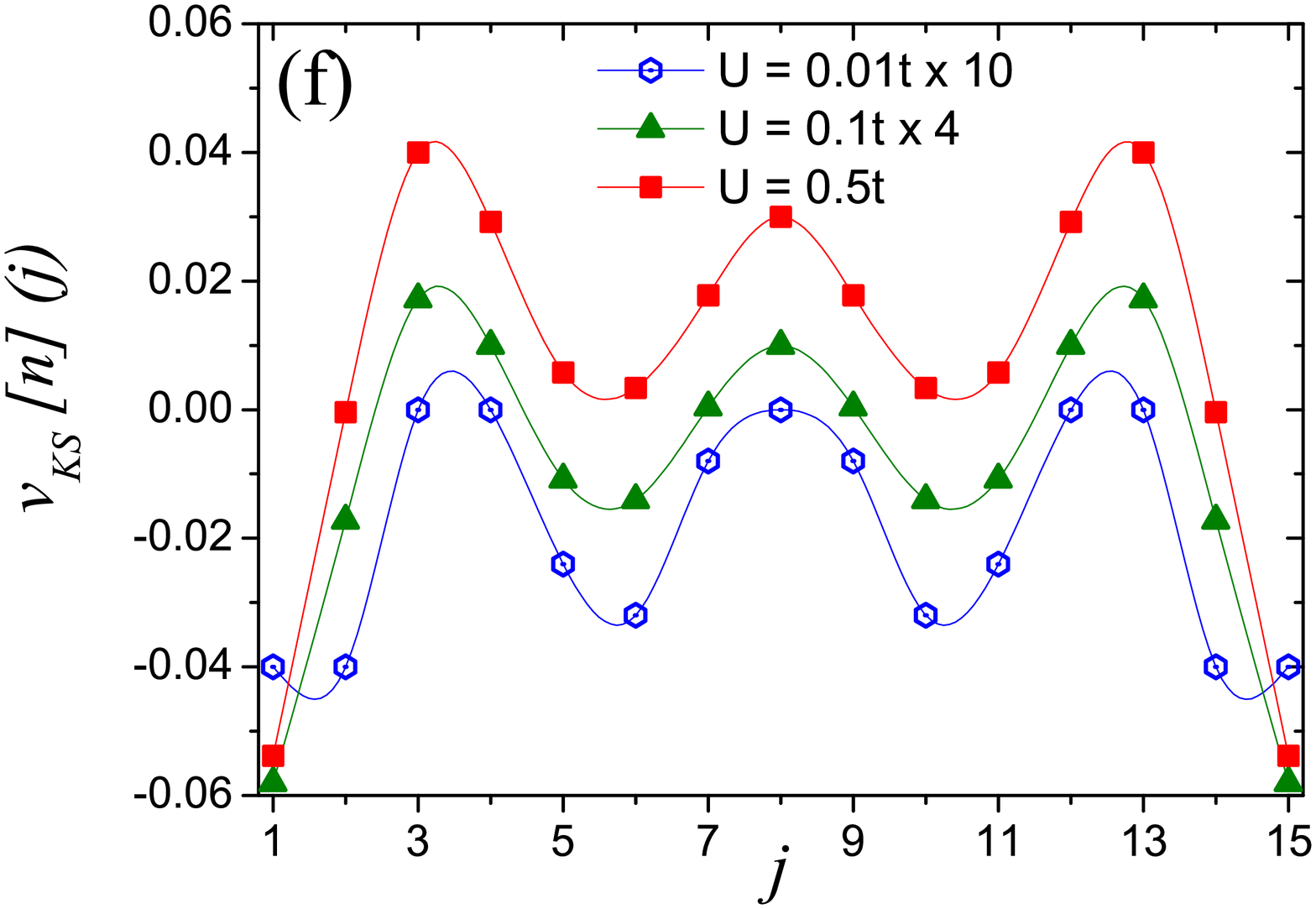} 
\end{minipage}
\caption{
\label{fig2} (Color online) Panels (a)-(b) and (e)-(f): Accurate (on Lanczos exact diagonalization accuracy) XC and KS potentials obtained by means of the KS equation inversion: $L=15$ sites and $N=6$ electrons, with $N_\uparrow = N_\downarrow$. Panels (c) and (d): the same as before, but on DMRG accuracy for $L=100$ sites and $N=16$ electrons. The average potentials were added to constant values in order to clarify the visualization.}
\end{figure*}

Using accurate density profiles as input, we numerically traced back the KS and XC potentials. The inversion of the KS equations is performed in a self-consistent cycle, as follows: {\it Step 1:} Introduction of a initial guess for the entire KS potential $V_{\textrm{KS}}^0(j)$, for example, which contains only the Hartree term of a uniform system. {\it Step 2:} Determine the density profiles $n^{0}(j)$ obtained from $V_{\textrm{KS}}^0(j)$. {\it Step3:} Compare $n^{0}(j)$ with the accurate density profiles $n^{\textrm{Accurate}}(j)$. {\it Step 4:} If $n^{0}(j) < n^{\textrm{Accurate}}(j)$, $V_{\textrm{KS}}^1 (j) = V_{\textrm{KS}}^0 (j) - \delta$; if $n^{0}(j) > n^{\textrm{Accurate}}(j)$,   $V_{\textrm{KS}}^1 (j)= V_{\textrm{KS}}^0 (j)+ \delta$, with $\delta \lesssim 1 \times 10^{-5}$. The $2 \rightarrow 4$ cycle is repeated $\gamma$ times until $n^{\gamma} (j) \approx n^{\textrm{Accurate}}(j)$, with an accuracy of $\sim 0.1\%$. Results are shown in Fig. \ref{fig2}, where we plot the accurate XC and KS potentials resulting from the density profiles of Fig. \ref{fig1} and from additional data. 

The answer about the v-representality is positive, that is, it is possible to find  spin-independent XC potentials which yield the accurate density profiles and their crossover in frequency. The same holds for the entire KS potential, $v_{\textrm{KS}}(j) = v_H[n](j) +v_{xc}[n](j) + v_{ext}(j)$, with $v_H[n](j)=Un(j)/2$ and here $v_{ext}(j)=0$. The comparison between Figs.~\ref{fig1} and \ref{fig2}(a)-(d) shows that, even though a change in  frequency is present for the densities, the XC potentials {\it always} oscillate with the same frequency, $4k_F$ (identified by the six or sixteen negative peaks, for each number of electrons). The only changes are in the oscillation amplitudes, which are increased as the interaction $U$ goes in the same direction. This behavior is specially unexpected for weak interactions, where the density profiles oscillate with $2k_F$, whereas the XC potentials do not follow the same path.
As consequence of an increment in the XC amplitude, the entire KS potential suffers a transition from $2 k_F$ to $4 k_F$, driving the same path to the density, as seen from Fig.~\ref{fig1}. Exceptions are some cases of very weak interactions, as pointed out in Figs. \ref{fig2}(e) and (f), where the XC potentials may oscillate with an intermediate frequency $2k_F < f < 4k_F$, identified by five negative peaks, and with no effects to the densities (which oscillate with $2k_F$ for these cases, due to the very reduced XC amplitudes).

As a first conclusion, the key ingredient which yields the $2k_F\rightarrow4k_F$ crossover in the density profiles is the change in amplitude of a $4 k_F$ oscillating XC potential. With the prominence of correlation effects, the spin-charge separation emerges from the XC potential, which oscillates with the same frequency as spinless electrons do.
 
\subsection{Approximated exchange-correlation potentials}

Once it is v-representable, the next option concerns the possibility of finding an approximated XC potential which yields the $2k_F\rightarrow4k_F$ crossover. The basic Hohenberg-Kohn and Kohn-Sham theorems of DFT also hold for the Hubbard model, once the density $n\r$ is replaced by the on-site occupation
number\cite{gsn}
\be
n\r   \longrightarrow n(j) 
= \sum_{\sigma=\uparrow,\downarrow} \la c_{j \sigma}^{\dagger}c_{j \sigma}\ra.
\ee 
In terms of this variable, local-(spin)-density approximations for
Hubbard chains and rings have been constructed. Successfull examples are the analytical parametrization proposed by Lima {\it et al.},\cite{balda} which is now called BALDA/LSOC, or its alternative route, which uses a direct numerical solution of the Bethe Ansatz integral equations, and is known as BALDA/FN (fully numerical).\cite{baldafn} Both approaches can be employed in the usual way, via a KS calculation.
  
\begin{figure*}
\centering
\includegraphics[scale=0.7]{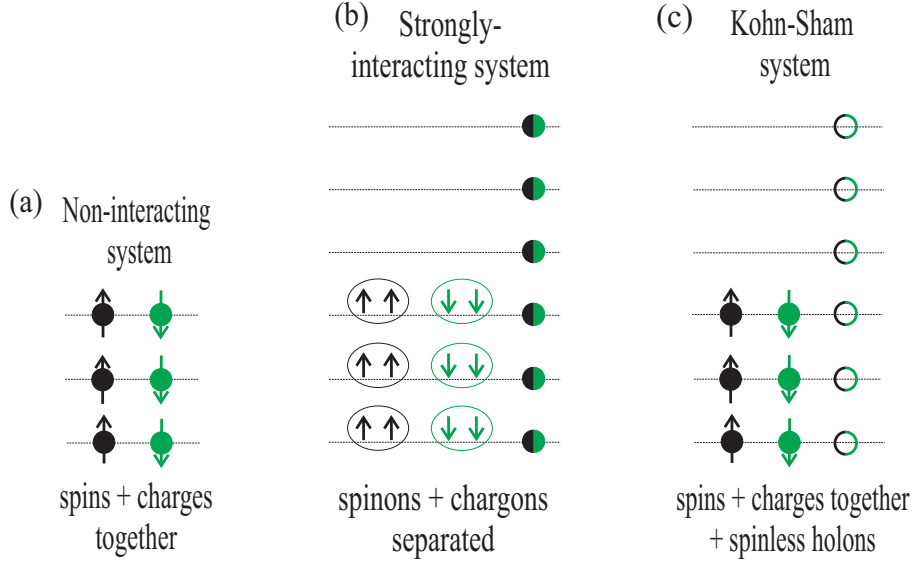}
\caption{
\label{fig3} (Color online) Schematic occupied states of three systems: (a) Non-interacting, with spins and charges together. (b) Strongly-interacting, with  spinons and chargons following the fractional statistics of semions. (c) Non-interacting Kohn-Sham system, with spins and charges together under influence of holons (the chargon antiparticles).}
\end{figure*}

The LDA functional, with $n_\uparrow(j) = n_{\downarrow}(j)$, is not seen to describe the Friedel oscillations correctly, at least in the limit of Wigner crystallization. We here will not consider the application of any vanishing magnetic field, which can break the spin symmetry and yield the $4k_F$ oscillations, however, at expense of a magnetized final solution with $n_\uparrow(j)\neq n_{\downarrow}(j)$ for $N_\uparrow = N_\downarrow$. Instead, we  propose a spin-independent XC potential, based on a complement of the previous BALDA approaches. The total Hamiltonian of an one-dimensional interacting system is known to separate into two independent terms, of spin and charge, as follows:\cite{giamarchi}
\be
\label{toth}
\hat{H} = \hat{H}_0 +  \hat{H}_\beta + \hat{H}_\rho = \hat{H}_0 + \hat{H}_I, 
\ee
where $\hat{H}_0$, $\hat{H}_\beta$ and $\hat{H}_\rho$ stand for the kinetic, spin and charge terms, respectively. Here, the spin densities $\beta(j)$ are built from  uncharged spin$-1/2$ electrons (spinons), whereas the charge densities $\rho(j)$ are built from charged spinless electrons (chargons). Spinons and chargons are semions, that is, particles with statistics half that of regular fermions,\cite{semions1, semions2,semions3} with occupied states following the schematic representation of Fig.~\ref{fig3}(b), where charge and spin were fractionalized to form the spin$-1/2$ spinons (as a sum of two $1/4$ spins) and the spinless chargons (as a sum of two $-e/2$ charges). In this context, considering spinons and chargons  as distinct separated entities, the total interacting density can be written as $n(j) = \beta(j) + \rho(j)$, with spin and charge allowed to travel at different velocities. Particularly, in the strongly-interacting limit, it has been shown that in some cases the spin densities can be considered as almost static in comparison with the charge moviment.\cite{schulz}  

\begin{figure*}
\centering
\begin{minipage}[b]{0.50\linewidth}
\includegraphics[scale=0.3]{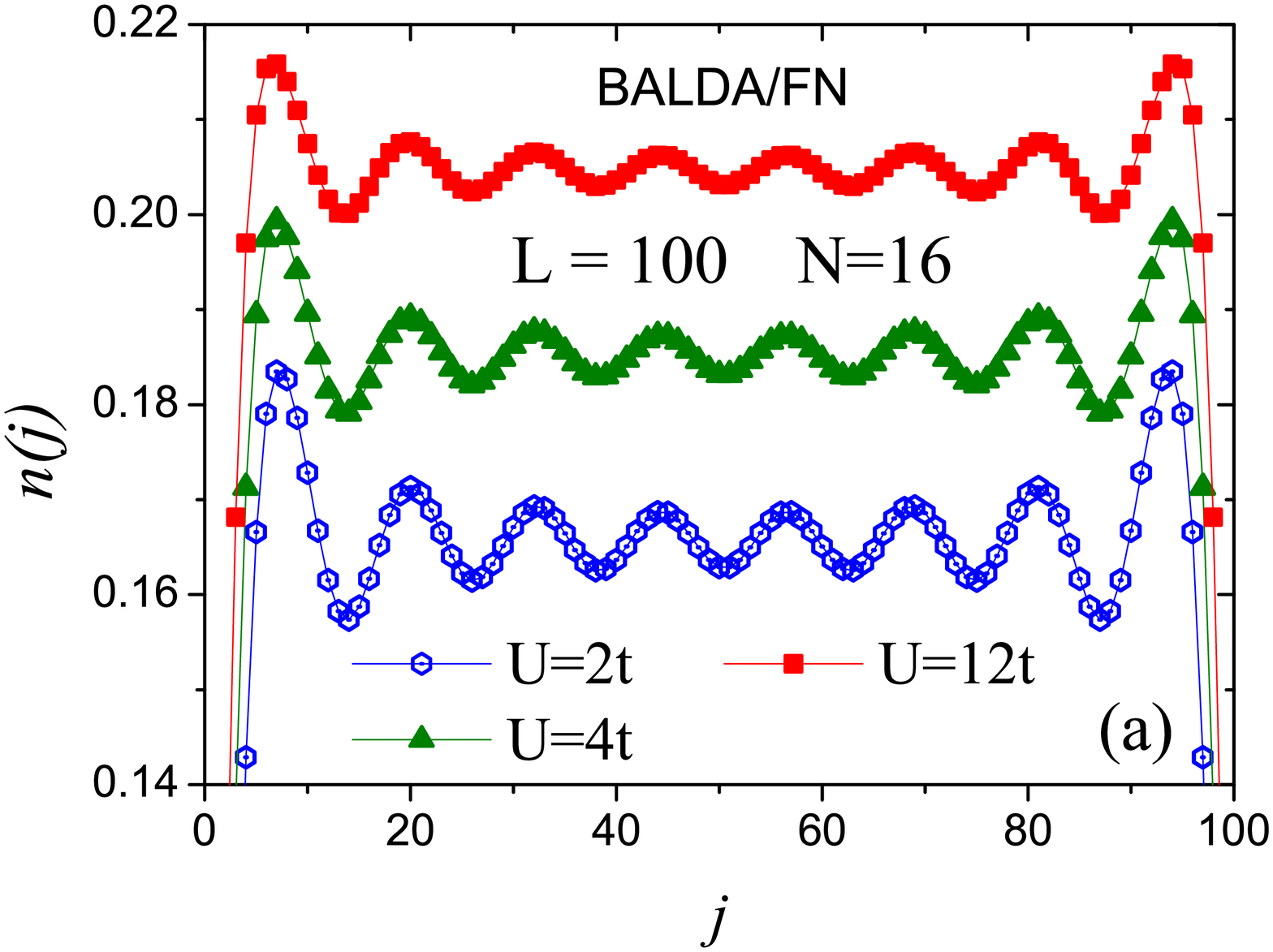}
\end{minipage}\hfill
\begin{minipage}[b]{0.50\linewidth}
\includegraphics[scale=0.3]{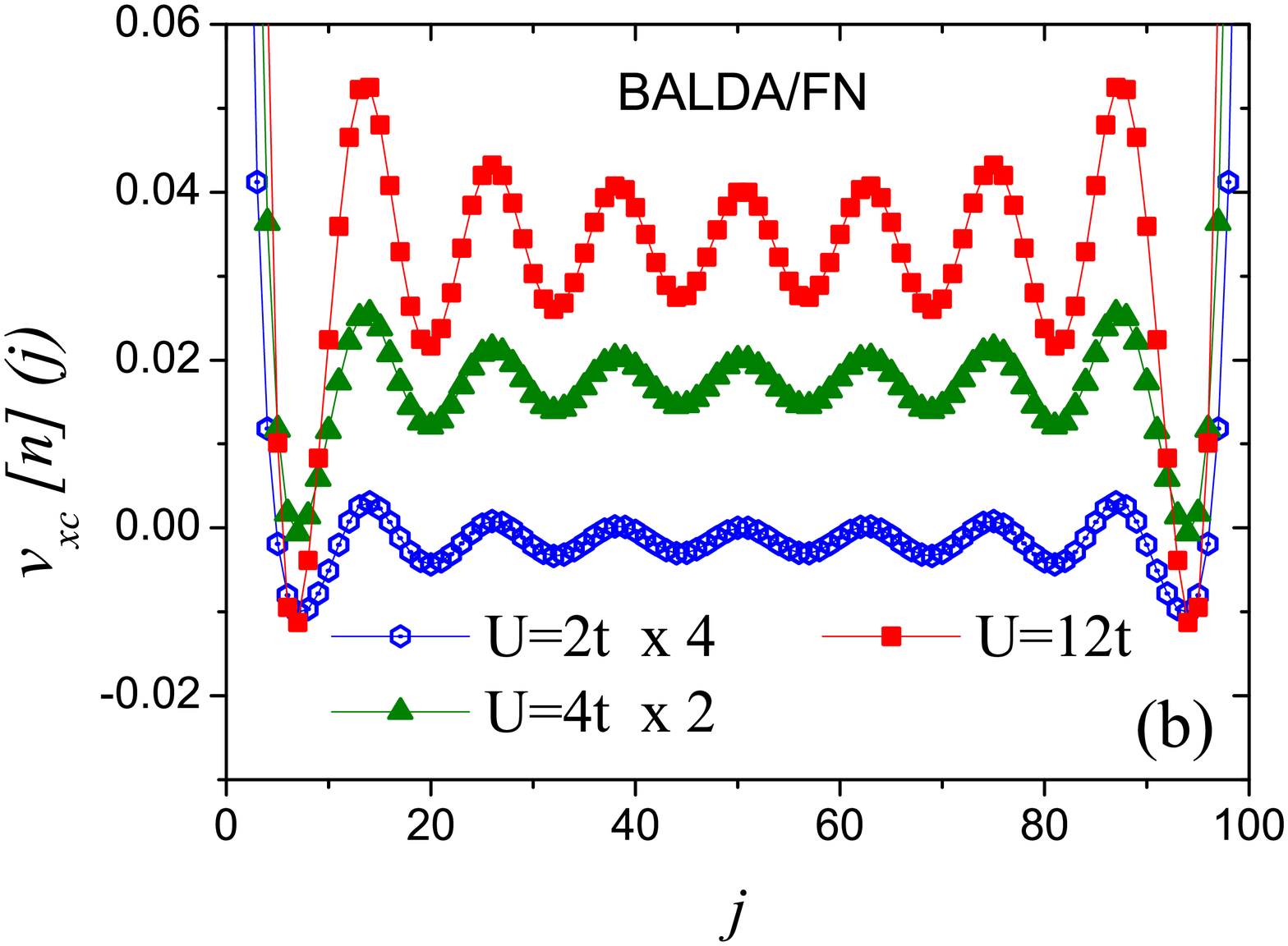}
\end{minipage}
\begin{minipage}[b]{0.50\linewidth}
\includegraphics[scale=0.3]{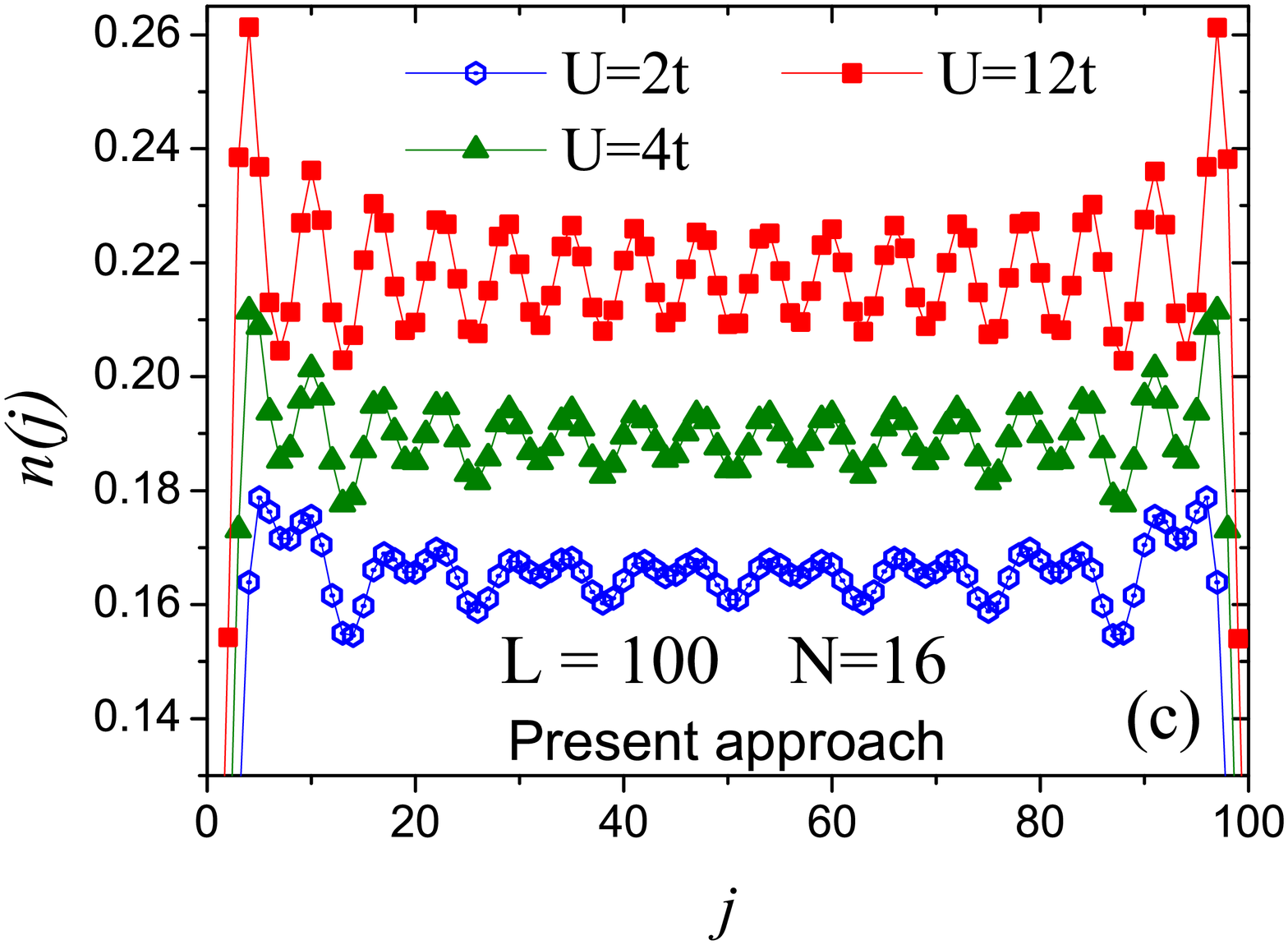}
\end{minipage}\hfill
\begin{minipage}[b]{0.50\linewidth}
\includegraphics[scale=0.3]{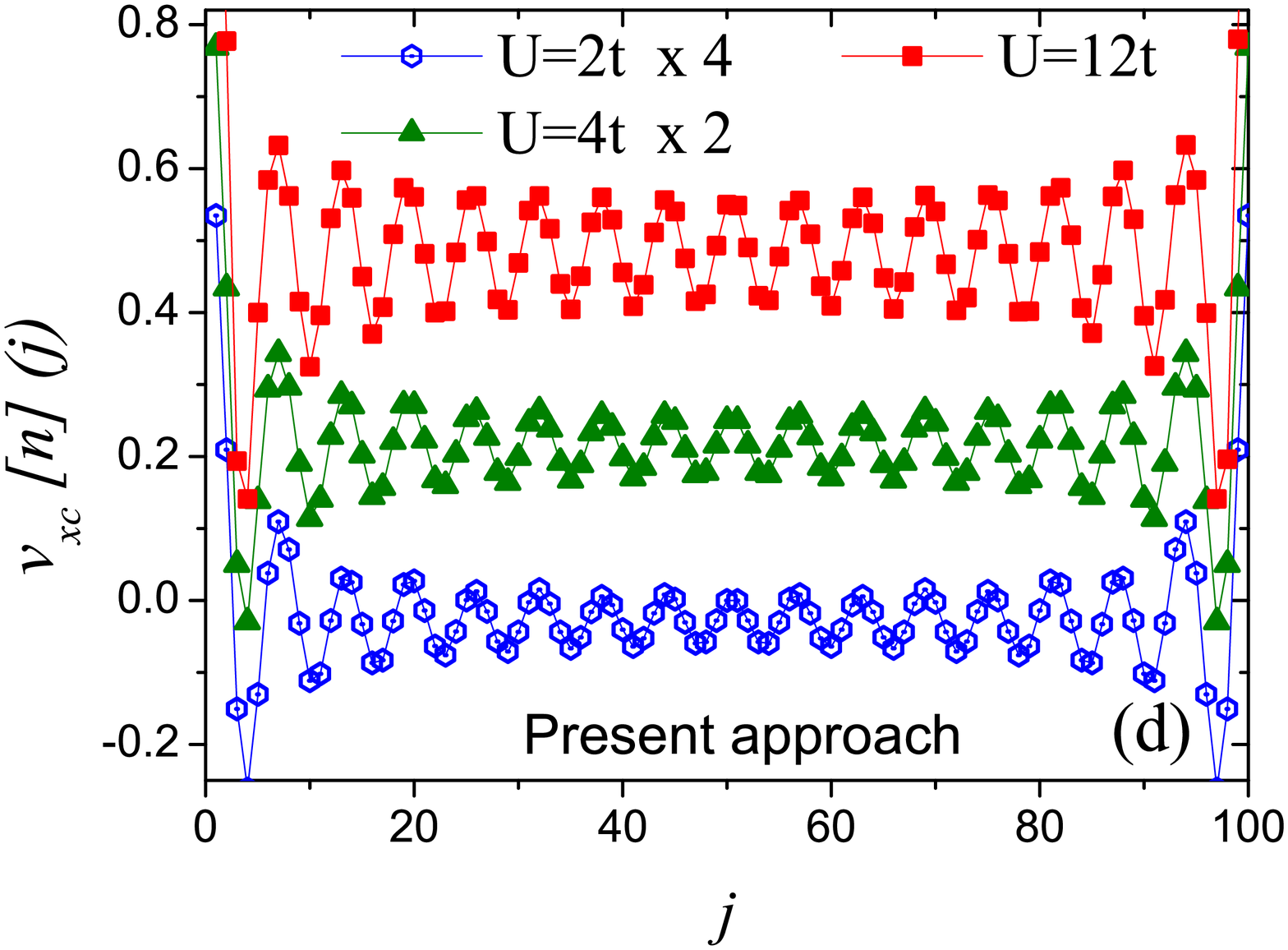}
\end{minipage}
\caption{(Color online) Panels (a) and (b): BALDA/FN density and XC potential profiles. Panels (c) and (d): Present approach profiles. $L=100$ sites and $N=16$ electrons, with $N_\uparrow = N_\downarrow$. }
\label{fig4}
\end{figure*}

\begin{figure*}
\centering
\begin{minipage}[b]{0.50\linewidth}
\includegraphics[scale=0.3]{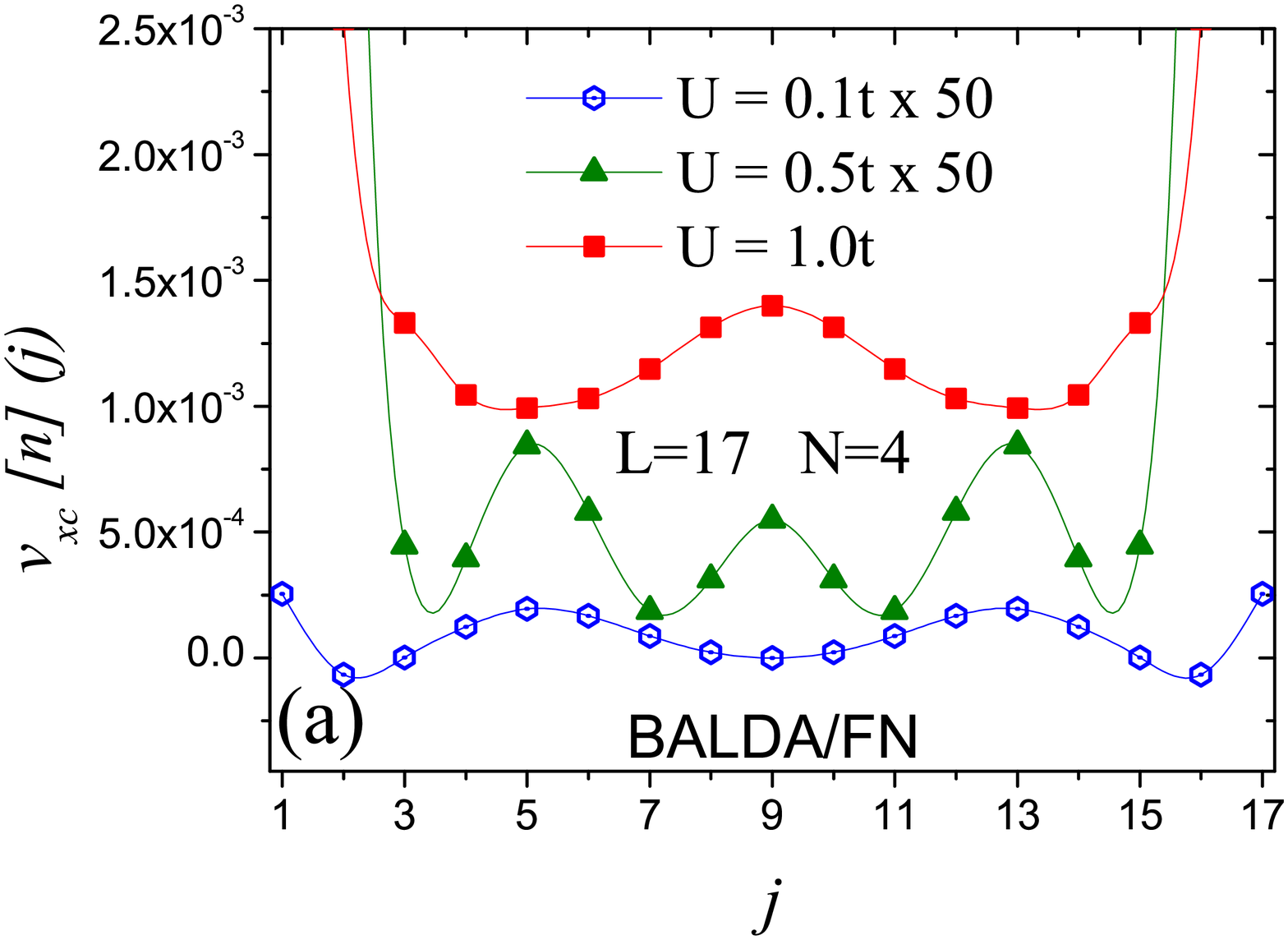}
\end{minipage}\hfill
\begin{minipage}[b]{0.50\linewidth}
\includegraphics[scale=0.3]{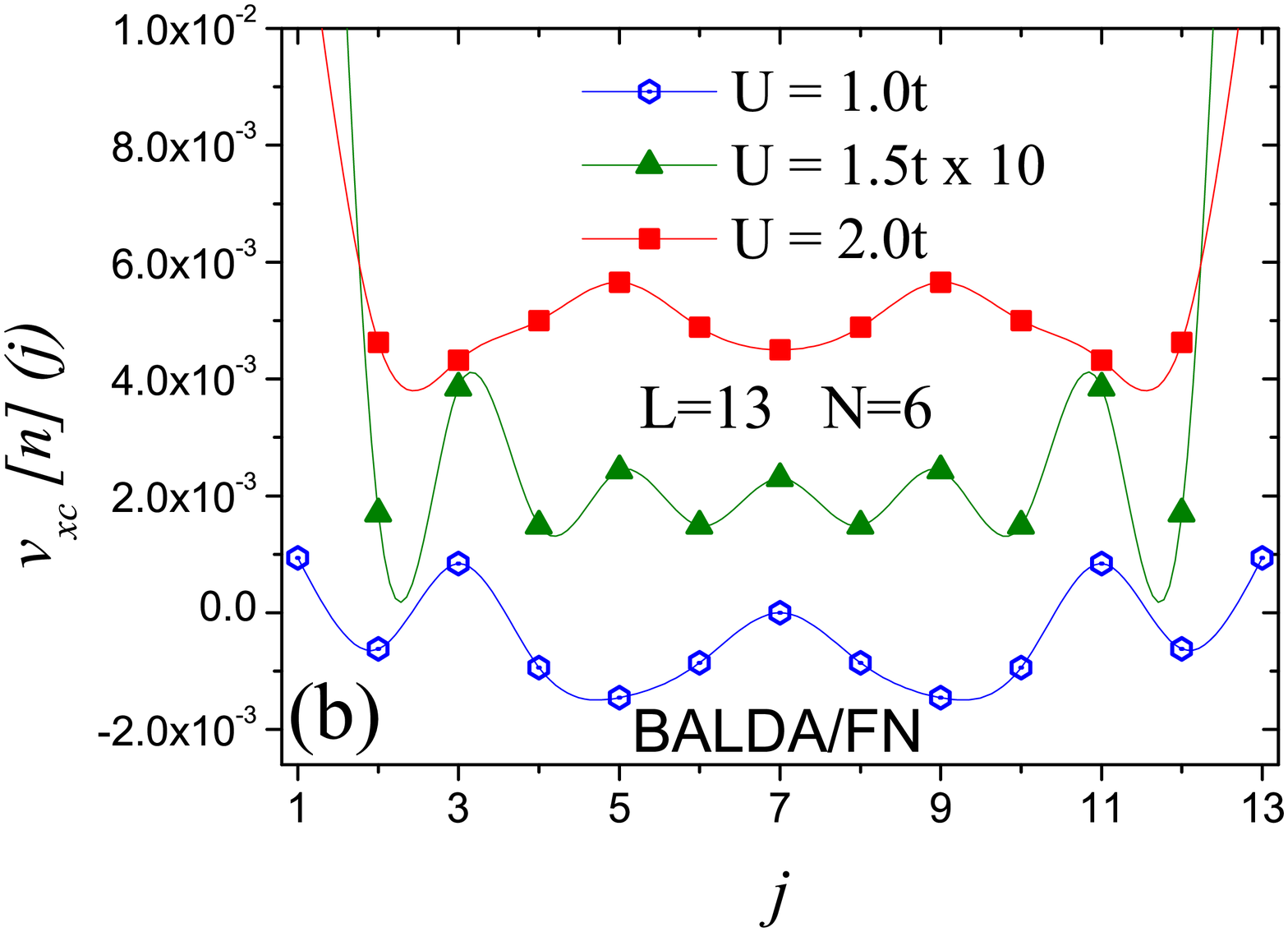}
\end{minipage}
\caption{(Color online) BALDA/FN XC potential profiles: (a) $L=17$ sites and $N=4$ electrons; (b) $L=13$ and $N=6$. Both cases with $N_\uparrow = N_\downarrow$.}
\label{fig5}
\end{figure*}

In contrast to spin$-1/2$ electrons and spin$-1/2$ holes in the Fermi-liquid theory, excitations in one-dimensional systems are described in terms of spinons and positively charged spinless holes -- the holons (the chargon antiparticles). It has been argued that spinons and holons attract each other at short distances.\cite{spinonholonattraction} In this context, as shown in Fig.~\ref{fig3}(c), we here propose that the occupied states of a non-interacting KS system are built by retaining spin and charge together, at expense of the presence of holons, whose densities are given by $\rho^+(j)$. The  KS potential can thus be written as a functional which depends on $n^{KS}(j)$ and $\rho^+(j)$:
\bea 
\label{totvks}
v_{\textrm{KS}}[n](j) &=&v_{ext}(j)+     \frac{\delta \la H_I \ra}{\delta n^{KS}(j)} -\frac{\delta \la H_I \ra}{\delta \rho^+(j)},
\eea
with $n^{KS}(j) \equiv n(j)$ and
\be 
\label{spinless}
\rho^+(j) =   \sum_{\sigma=\uparrow,\downarrow} \sum_{k}^{N}\ \frac{1}{2}\ |\psi_{k\sigma }(j)|^2, 
\ee
that is, the holon density $\rho^+(j)$ is constructed by using the first $N_{{\sigma}}$ unoccupied KS orbitals.
This conjecture is specially conceived to deal with strong interactions,
where the presence of holons will be felt by the electronic density as an attractive external potential. Based on Eqs.~(\ref{totvks}) and (\ref{spinless}), we  introduce the present approach:
\bea
\label{present}
v_{\textrm{KS}}^{\textrm{Present}}[n](j)&=&v_{ext}(j) + v_H[n](j) +v_{xc}^{approx}[n](j)
\nonumber \\
 & &\ \ \ \ \ \ \ \ \ - v_H[\rho^+](j)-v_{xc}^{approx}[\rho^+](j) \nonumber
\\
&\equiv &v_{ext}(j) + v_H[n](j) + v_{xc}^{\textrm{Present}}[n](j),
\label{pzsicV2}  
\eea
with
\be 
n(j) =  \sum_{\sigma=\uparrow,\downarrow} \sum_{k}^{N_\sigma}\   f_{k,\sigma}\ |\psi_{k\sigma }(j)|^2,
\ee
and the KS choices for $f_{k,\sigma}$. As approximated functional ({\it approx}) we chose the BALDA/FN. We include the holon density only in the XC potential, and use it in the KS calculation. A welcome feature of the present approach is the absence of the one-electron self-interaction error: In a non-magnetized LDA calculation, $\psi_{k\uparrow }(j)=\psi_{k\downarrow}(j)$. Therefore, for one-electron systems, $n(j) \equiv \rho^+(j) =  |\psi_{k\sigma }(j)|^2$, and then the present KS potential is exact, correcting the spurious self-interaction of one electron with itself. The results for BALDA/FN and the present correction are shown in Figs. \ref{fig4} and \ref{fig5}, for the electronic density $n(j)$ and XC potential. Remember: BALDA/FN takes into account only the first  two terms of Eq. (\ref{totvks}).

From Fig.~\ref{fig4}(a), while BALDA/FN yields the correct behavior for weak interaction, the $2k_F\rightarrow4k_F$ crossover is not recovered as $U$ is increased, following the same trends already observed in the literature.\cite{vieira,xianlong0,xianlong1,xianlong2} The same occurs with the BALDA/FN XC potentials of Fig.~\ref{fig4}(b), which oscillate in $2k_F$ for all values of $U$ (identified by the eight negative peaks). From Fig.~\ref{fig4}(c), the densities from the present approach also \mbox{oscillate} correctly for weak interactions, but, different from BALDA/FN, we do observe the  $2k_F\rightarrow4k_F$ crossover as  $U$ is increased, even though the amplitudes are subtly higher than the DMRG prediction of Fig.~\ref{fig1}. In addition, from Fig.~\ref{fig4}(d), the XC potentials from the present approach oscillate in $4k_F$ for all values of $U$, the only changes are in the amplitudes, the same as observed for the accurate XC profiles of Fig.~\ref{fig2}. The spinless term of Eq.~(\ref{totvks}) was able to incorporate the key ingredient we mentioned before, {\it i.e.}, the spin-charge separation by means of a change in amplitude of a $4k_F$ oscillating XC potential.

\begin{figure*}
\centering
\begin{minipage}[b]{0.50\linewidth}
\includegraphics[scale=0.28]{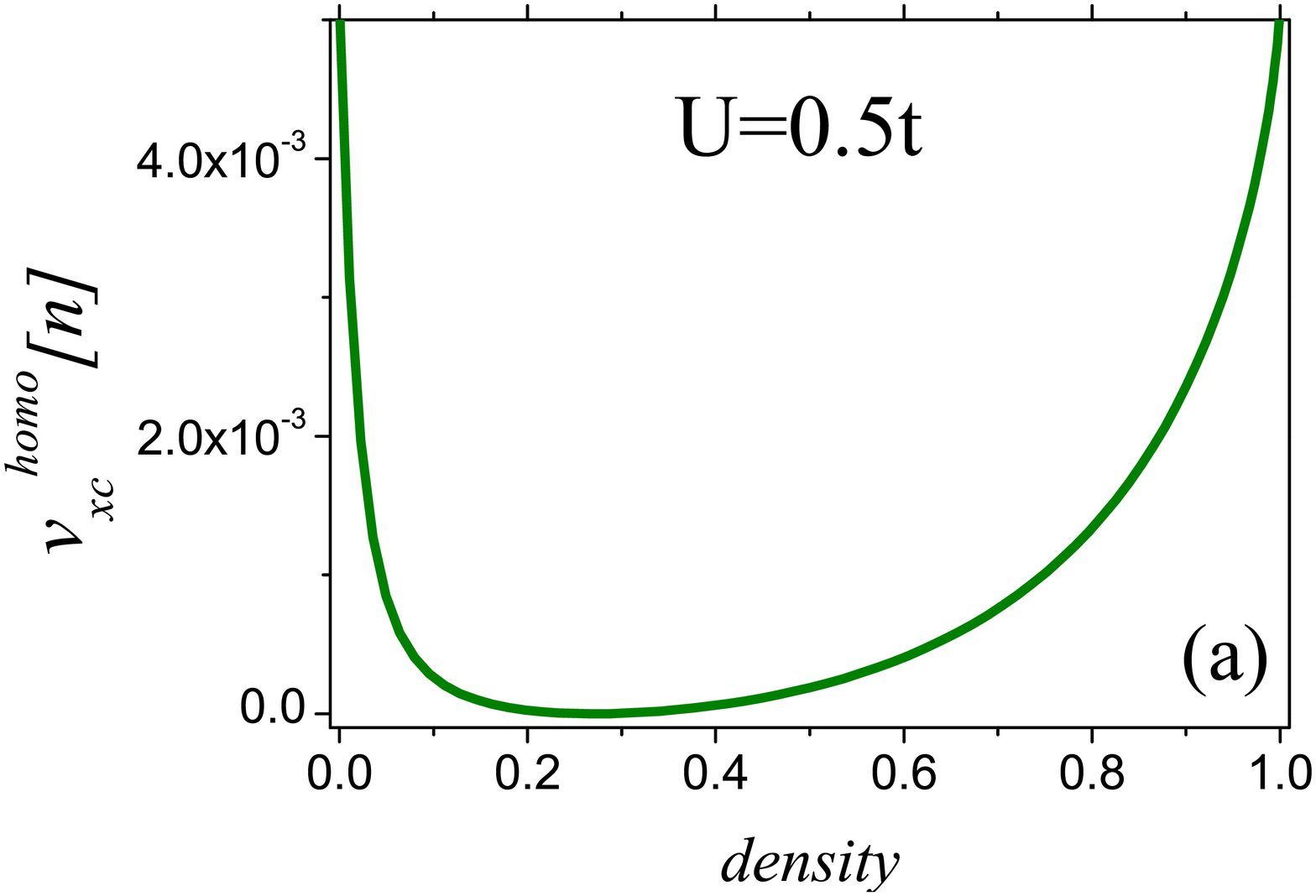}
\end{minipage}\hfill
\begin{minipage}[b]{0.50\linewidth}
\includegraphics[scale=0.28]{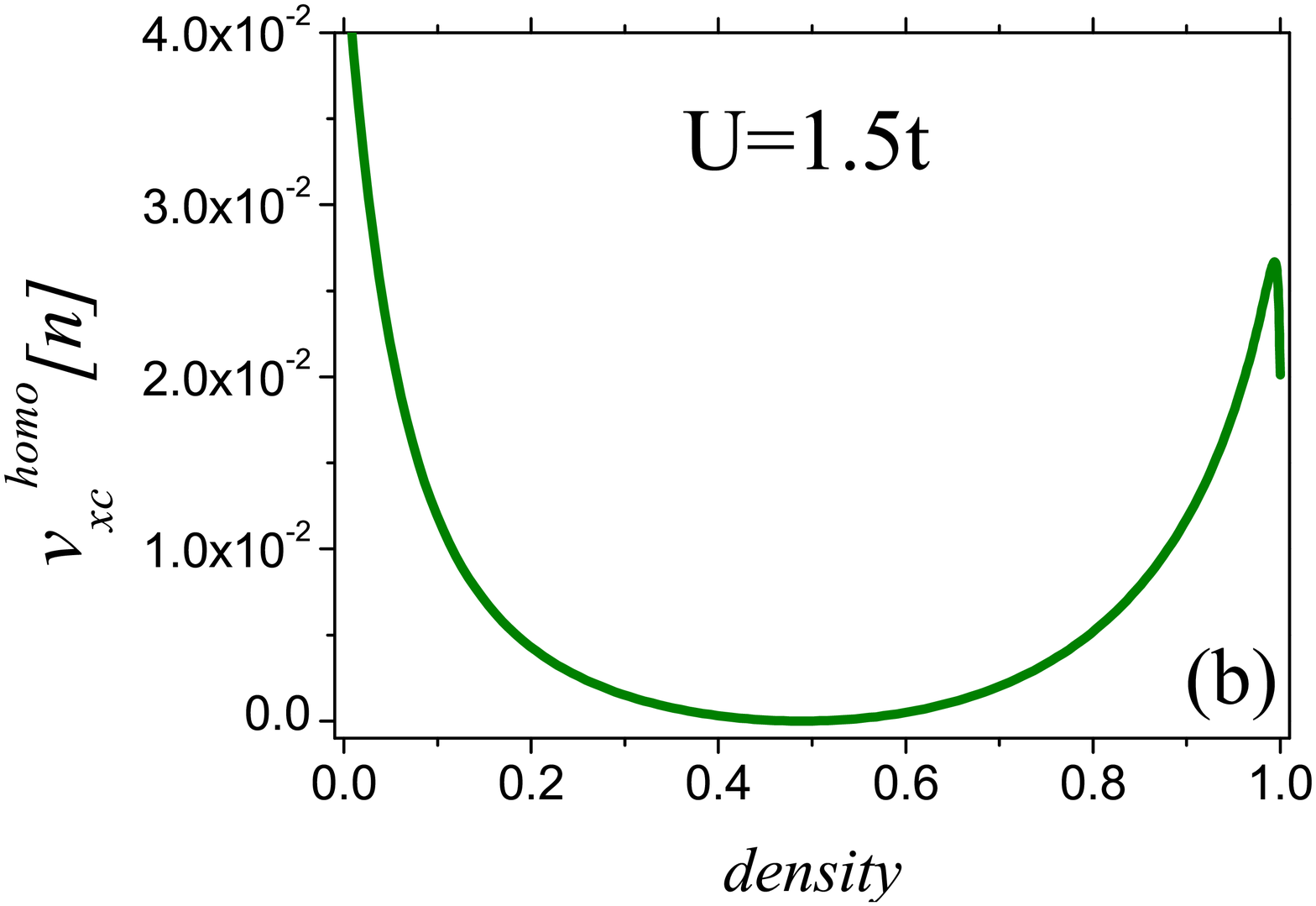}
\end{minipage}\hfill
\begin{minipage}[b]{0.50\linewidth}
\includegraphics[scale=0.28]{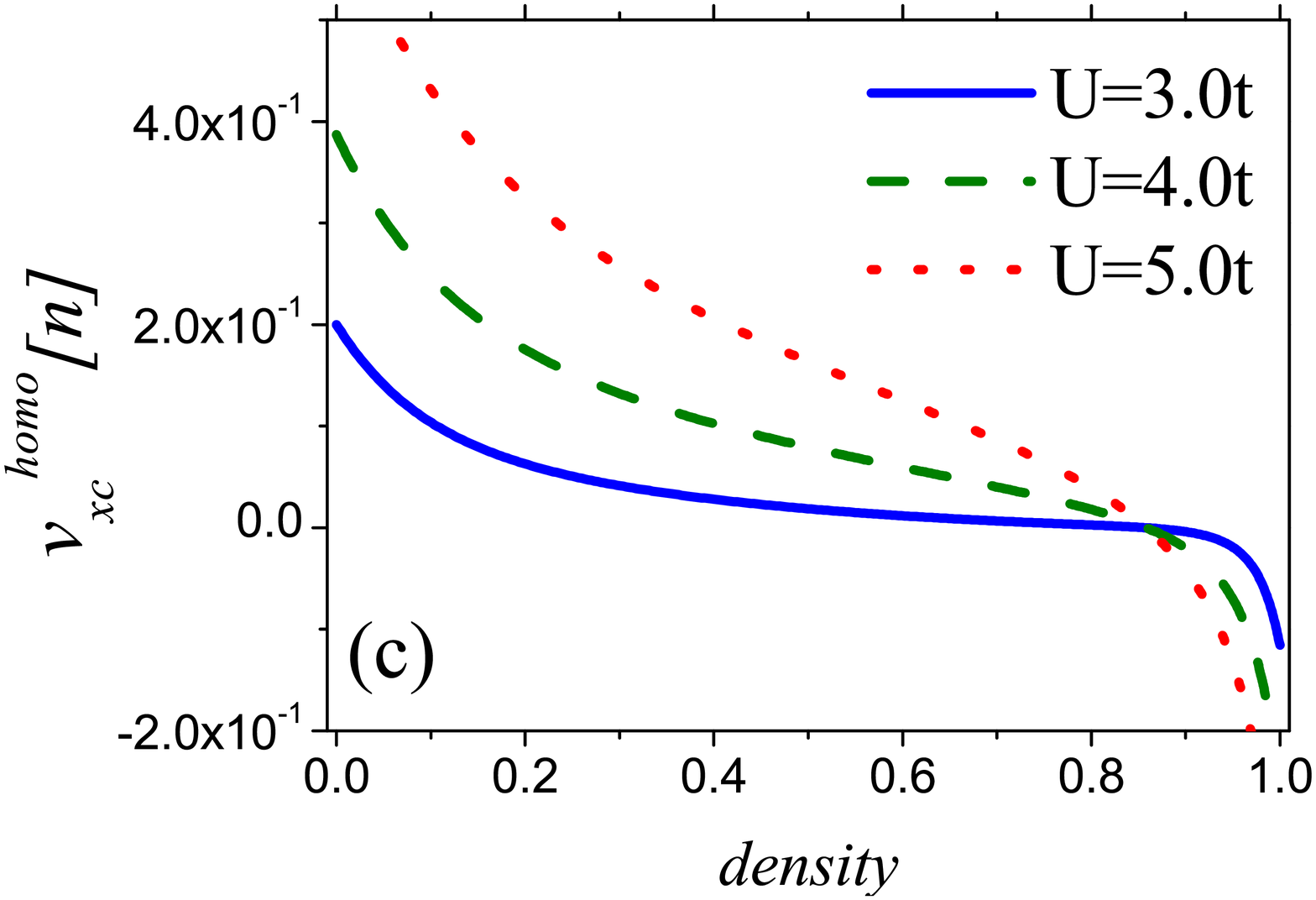}
\end{minipage}\hfill
\begin{minipage}[b]{0.50\linewidth}
\includegraphics[scale=0.28]{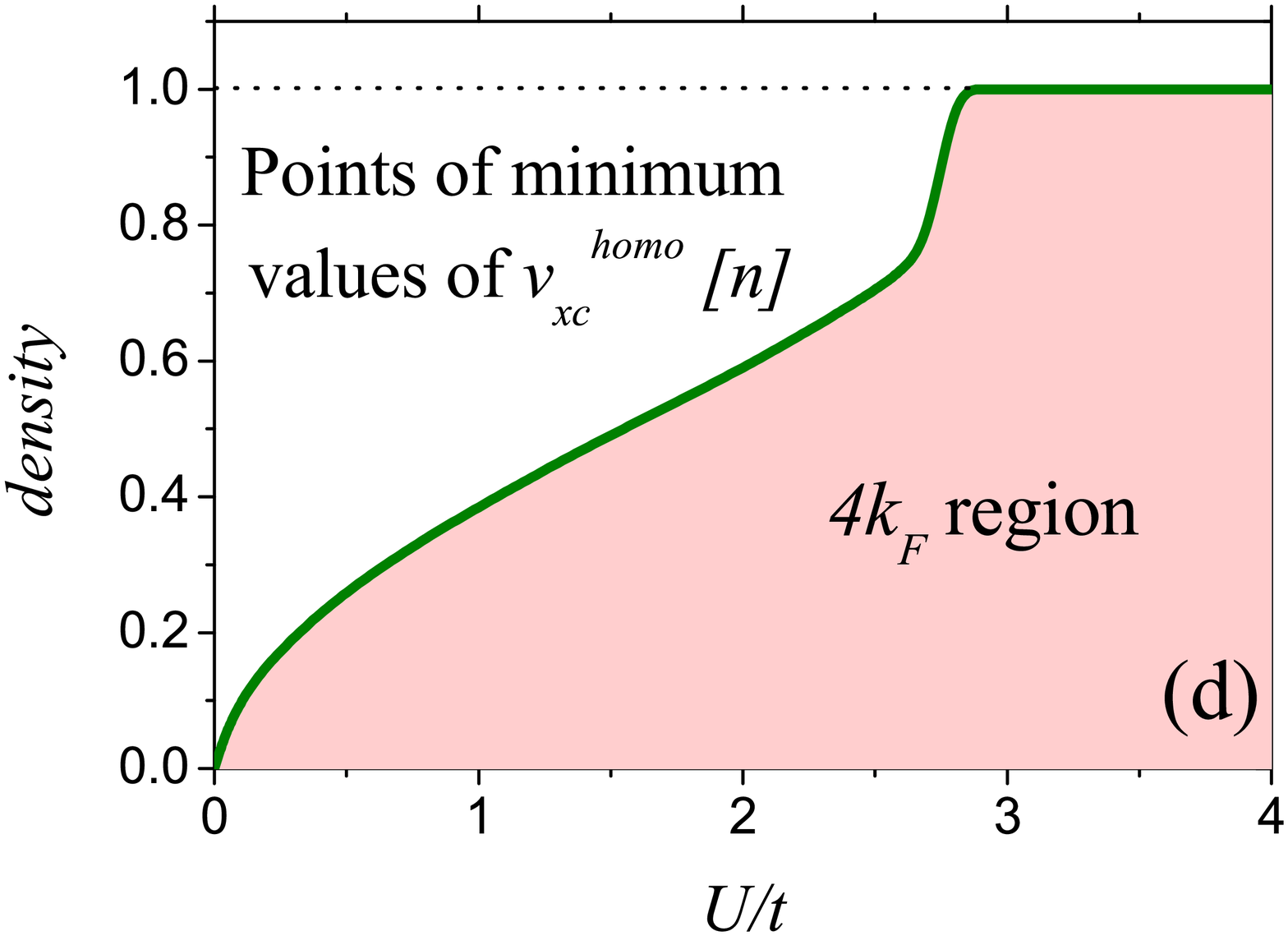}
\end{minipage}
\caption{(Color online) Panes (a)-(c): XC potential profiles for extended homogeneous systems obtained by means of the Bethe Ansatz integral equations. Panel (d): Values of $n$ for which the homogeneous XC potential undergoes a minimum -- and the LDA XC potential oscillates with $4k_F$ (for $n<1$).}
\label{fig6}
\end{figure*}

It is known that the prominence of correlation effects is not only driven by the interaction $U$, but also by low densities. For this reason, we have also considered Hubbard chains with different average densities $n=N/L$. Specifically, in Figs.~\ref{fig5}(a)-(b) we plot the LDA XC potentials for two values of $n$. Interestingly, for each case, there are values of $U$ for which the LDA XC potentials oscillate in $4k_F$, as signed by the four (when $N=4$) or six (when $N=6$) negative peaks. As $U$ is increased, the LDA XC potential suffers an incorrect transition to $2k_F$ (identified by two or three negative peaks), and then yields the incorrect density profiles for larger $U$ (not shown). This feature can be better understood by means of the XC potentials of homogenous extended systems, which are used in the construction of the LDA XC functionals, and are plotted in Figs.~\ref{fig6}(a)-(c). We note a common behavior: Near the average densities of the systems of Fig. \ref{fig5}, the homogeneous XC potentials undergo minimum values, whose positions tend to move to the right as the interaction is increased. For a given value of $U$, the LDA XC potential emerges to oscillate in $4k_F$  only for the average densities surrounding these minimum values. Without the inversion points, as seen for $U\gtrsim 3t$ in Fig.~\ref{fig6}(c), an increase of $n(j)$ will be always followed by a decrease of $v_{xc}[n](j)$, and vice-versa, explaining correspondence between Figs.~\ref{fig4}(a) and (b). In Fig. \ref{fig6}(d) we plot the values of $n<1$ for which the LDA XC potentials oscillate with $4k_F$: For average densities above this curve, the LDA XC potentials oscillate with an intermediate frequency $2k_F < f < 4k_F$, as identified by three and four negative peaks in Figs. \ref{fig5}(a) and (b), respectively, in accordance with the similar behavior of Fig. \ref{fig2}(e). Bellow the curve, the LDA XC potentials oscillate with the wrong frequency $2k_F$. 

To conclude, {\it (i)} it should be stressed that even though the LDA XC potential may oscillate with $f$ and $4k_F$, these frequencies occur for values of $U$ in which the XC amplitudes are so reduced that the density profiles are not affected, retaining the $2k_F$ oscillations in all cases we have implemented; {\it (ii)} the curve of Fig.~\ref{fig6}(d) can be also employed to mark the necessity of using the KS states exposed in Fig.~\ref{fig3}(c), where the spin-charge separation emerges from the inclusion of spinless holons. As expected, the separation is prominent both for strong interactions and/or low densities, situation indicated in Fig.~\ref{fig6}(d) as the ``$4k_F$ region''. 

The present approach yields the correct frequencies of density profiles for all values of $U$ and $n$ we have investigated, including the systems of Fig. \ref{fig5} (not shown). This is another numerical proof that is  possible to find a spin-independent KS potential which yields the Friedel and Wigner crystal oscillations, for all ranges of interaction and prominence of correlation effects. All this in the following implementation scenario (on a supercomputer): The computing time of DMRG scales with $\sim 10^1$ hours, while the present DFT approach with $\sim  10^0$ minutes. We have also considered other flavors of self-interaction corrections, such as the full Optimized Effective Potential (OEP) implementation of the usual Lundin-Eriksson\cite{lesic} and Perdew-Zunger\cite{pzsic}  \mbox{approaches}. However, none of them were able to yield the $4k_F$ oscillations, what teach us that the $2k_F\rightarrow4k_F$ crossover, or the spin-charge separation, is not only a problem of {\it one-electron} self-interaction error.\cite{1sie} More than correcting the spurious self-interaction of LDA, which tends to delocalize electrons, an accurate functional should include the physics of positively charged holons.

\subsection{Analytical conjecture}
\label{accurate}

Based on Eqs.~(\ref{eqfriedel1}) and (\ref{eqfriedel2}), it is possible to propose an analytical conjecture to describe the Friedel and Wigner crystal oscillations, valid for any value of $U$. It is usually written as:\cite{wignerprb} 
\be
\label{eqfriedel} 
n(j) = \frac{N+1}{(L+1)} - A_1 \frac{\sin(2k_F j)}{\left[ \sin \left (\frac{\pi j}{L+1} \right)\right]^{(K_c+1)/2}} 
 -A_2 \frac{\sin\left[(4k_F  - \frac{\pi}{L+1})j \right]}{\left[ \sin \left (\frac{\pi j}{L+1} \right)\right]^{2K_c}},
\ee
The parameters $A_1$, $A_2$ and $K_c$ depend on $U$ and are related to the amplitudes and oscillations decay. Specifically, $A_1$ and $A_2$ are usually referred to as Friedel and Wigner amplitudes, respectively. $K_c$ is the correlation exponent, one of the  Luttinger parameters.\cite{schulz} A similar expression has  been proposed for the density distribution in a harmonic trapping potential.\cite{wignerpra} We here fitted the density profiles of Figs. \ref{fig1} (b) and \ref{fig4} to the expression (\ref{eqfriedel}), with the values of $A_1$, $A_2$ and $K_c$ expressed in Table \ref{sictable} as a function of $U$. In comparison with DMRG, LDA fails to predict the correct values of $A_2$, as well as, the correct decay of $A_1$ and $K_c$.  The present approach performs much better, mainly for the $K_c$ values, which are obtained in great agreement with DMRG, even though it is seen to overestimate the weight of $A_2$ for large $U$, as expected from the density profiles of Fig. \ref{fig4}(c).

\begin{figure}
\centering
\includegraphics[scale=0.35]{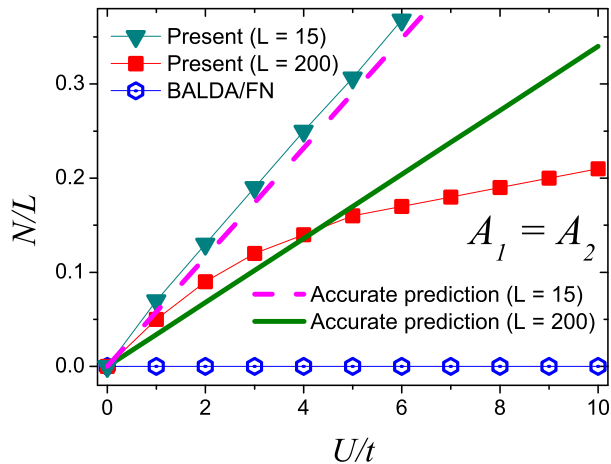}
\caption{
\label{fig7} (Color online) Friedel and Wigner crystal amplitudes. Crossover points with $A_1 = A_2$: For average densities above the curves, systems display Friedel-like oscillations. Bellow them, the Wigner crystal behavior is prominent. }
\end{figure}

Beyond the local interaction $U$, all parameters in Table \ref{sictable} depend on the number of particles $N$, chain size $L$ and average density $N/L$. For this reason, these values cannot be considered as universal -- they are only a picture of a specific system with $N=16$ and $L=100$. A more complete analysis follows in Fig. \ref{fig7}, where we plot the crossover points with $A_1 = A_2$ in the $U$ - $N/L$ plane. Based on DMRG calculations, an accurate prediction for $L=200$ is $N/L \approx 0.034\ U$.\cite{wignerprb} For a small chain of $L=15$ sites, we here introduce an additional accurate expression, $N/L \approx 0.058\ U$, based on Lanczos exact diagonalization. For average densities above these curves, systems display Friedel-like oscillations with $2k_F$ ($A_1>A_2$). Below it, the Wigner crystal behavior with $4k_F$ is prominent ($A_1<A_2$). The constant BALDA/FN curve at zero means that it predicts $A_1>A_2$ for all values of $U$ and $N/L$. The curves from the present approach emerge to be much more accurate, mainly for low densities and/or moderate interactions. Its great improvement over BALDA/FN is due to a more accurate description of the $2k_F \rightarrow 4k_F$ crossover. For strong interactions and higher average densities, as seen for $L=200$, the present approach tends to fit below the accurate prediction, yielding a larger Friedel-like region.
\begin{table*}  
\caption{\label{sictable} Parameters $A_1$, $A_2$ and $K_c$ obtained by means of a fit of expression (\ref{eqfriedel}) to the data of Figs. \ref{fig1} (b) and \ref{fig4}. The nomenclature ``x.xx / y.yy / z.zz'' indicates the parameters values for $U=2t$, $U=4t$ and $U=12t$, respectively.}
\begin{tabular}{c|c|c|c}
\hline \hline
$L=100$ and $N=16$ & $A_1 \times 10^{-3}$ & $A_2 \times 10^{-3}$& $K_c$   \\
\hline
DMRG &   5.63 / 3.60 / 1.57 &    1.34 / 2.53 / 3.90 &    0.65 / 0.58 / 0.53 \\
\hline
LDA   & 5.70 / 4.78 / 3.84 &  0.19 / 0.38 / 0.45  & 0.89 / 0.82 / 0.80 \\
\hline
Present    &5.25 / 2.52 / 0.36  & 2.00 / 3.35 / 4.54 & 0.61 / 0.55 / 0.52 \\ 
\hline \hline
\end{tabular}
\end{table*}

\section{Conclusions}
 
We have dealt with the v-representability issue of Wigner crystal oscillations in one-dimensional Hubbard chains. Initially, we posed two questions concerning non-magnetized systems: (1) Are the 1D Wigner crystal oscillations non-interacting v-representable? (2) Or, is there a spin-independent KS potential which is able to yield the spin-charge separation? We have concluded that the answers to both questions are positive, that is, we do not face a v-representability problem. Different from some previous studies, we have not considered the break of spin symmetry, which is known to yield the $2k_F\rightarrow4k_F$ crossover. Instead, by means of accurate many-body solutions for small and extended chains, we inverted the KS equation, showing that it is possible to find a spin-independent XC potential which reproduces the accurate densities. 

In addition, we have proposed a new XC potential, which incorporates, by means of unoccupied KS orbitals, the physics of positively charged spinless holes (holons). Specifically, holons act on mediating attractive interactions, as an additional contribution to the repulsive on-site $U$. The present approach is also able to reproduce the $2k_F\rightarrow4k_F$ crossover, either for small or extended chains, with correlation exponents obtained in great agreement with DMRG data.

It remains to be seen whether the present spin-independent XC potential (\ref{present}) can circumvent known failures of LDA,\cite{failures} as well as, be successfully applied to other classes of systems, with impurities and external confinements, as the example of ultracold fermions in one-dimensional traps.\cite{wignerpra} 

{\it Aknowledgments} The author thanks Andre Malvezzi for the DMRG data, and Vivaldo L. Campo for his code of exact diagonalization. The author also thanks K. Capelle for useful discussions.

\end{document}